\input{aipcheck}

\documentclass[
    ,final            
  ]
  {aipproc}

\layoutstyle{6x9}

\newcommand{\la}{$\Lambda$}
\newcommand{\al}{$\bar{\Lambda}$}
\newcommand{\lll}{$\Lambda(\bar{\Lambda})$}

\begin{document}

\title{
Longitudinal Spin Transfer in Inclusive  $\Lambda$ and $\bar \Lambda$ Production in Polarized Proton-proton Collisions at $\sqrt s$ =200 GeV}


\author{Qinghua Xu for the STAR collaboration}{
  address={MS 70R0319, Lawrence Berkeley National Laboratory, Berkeley, CA 94720  
}
}

\begin{abstract}
This contribution reports on a proof-of-principle measurement of the longitudinal spin transfer $D_{LL}$ in inclusive $\Lambda$ and $\bar \Lambda$ production in polarized proton-proton collisions at a center of mass energy $\sqrt{s}$ = 200 GeV.  
The data sample consists of about 3 $\times$ $10^6$ minimum bias events collected in the year 2005 by the STAR experiment at RHIC with proton beam polarizations of up to 50\%.
The $\Lambda$($\bar \Lambda$)  candidates are reconstructed at mid-rapidity ($|\eta|<1$) using the STAR Time Projection Chamber via the dominant decay channel $\Lambda\to p \pi^-$ ($\bar \Lambda \to \bar p \pi^+$). 
Their mean transverse momentum $p_T$ is about 1.3 GeV/c and longitudinal momentum fraction
$x_F = 7.5 \times 10^{-3}$ .
The longitudinal $\Lambda$($\bar \Lambda$) polarization is determined using a method in which the detector acceptance mostly cancels. 

\end{abstract}
\maketitle


One of the main goals of the STAR (Solenoid Tracker At RHIC) spin physics program is to delineate the flavor composition of the proton spin.  
Measurements of longitudinal spin transfer, referred to as $D_{LL}$, for inclusive $\Lambda$ and $\bar \Lambda$ at large transverse momenta may provide constraints on strange (anti-)quark polarization in the polarized nucleon~\cite{XLS05}. It can also yield new insights into polarized fragmentation functions~\cite{Florian98}.
The spin transfer $D_{LL}$ measures the transfer of longitudinal polarization from the polarized beam to the \lll\ in singly polarized proton collisions,
\begin{equation}
D_{LL}\equiv \frac
{\sigma^{pp_+ \to  \Lambda _+ X}-\sigma^{pp_+ \to  \Lambda _-X}}
{\sigma^{pp_+ \to  \Lambda _+ X}+\sigma^{pp_+ \to  \Lambda _-X}},
\label{gener1}
\end{equation}
where $\sigma$ denotes the inclusive production cross section and the subscripts $+$ and $-$ the helicity.
It thus is the polarization of the \lll\ along its moving direction in an experiment with a fully polarized proton beam.
We expect that $D_{LL}$ is about 4 times larger for positive pseudo-rapidities with respect to the polarized beam than the double longitudinal beam spin asymmetry $A_{LL}$ and, therefore, that $D_{LL}$ is more sensitive than $A_{LL}$ to strange (anti-)quark polarization in the polarized nucleon.

%
The polarization of the \lll~ can be determined  via the weak decay channel $\Lambda \to p \pi^-$ ($\bar \Lambda \to \bar p \pi^+$) from the angular distribution of the decay (anti-)proton in the \lll\ rest frame:
\begin{equation}
\frac{dN}{d \cos{\theta}}=\frac{N_{tot}}{2}A(\cos\theta) (1+\alpha P_{\Lambda(\bar \Lambda)}
\cos{\theta}), 
\label{ideal}
\end{equation}
where $N_{tot}$ is the total number of \lll 's,
$\alpha=+(-)0.642\pm0.013$ is \la(\al)~decay parameter~\cite{PDG},
$P_{\Lambda(\bar \Lambda)}$ is the polarization of \lll~, $\theta$ is the angle
between the (anti-)proton momentum in the \lll\ rest frame and the polarization vector of \lll, and $A(\cos{\theta})$ is the detector acceptance versus $\cos{\theta}$ averaged over other observables.
In general, the detector acceptance depends on observables such as \lll\ momentum and the primary vertices position in addtion to $\cos{\theta}$.

In this analysis, about 3$\times$$10^6$ minimum bias events are used.
They were collected in the year 2005 with the STAR experiment~\cite{nim} in proton collisions at $\sqrt s$ = 200 GeV with longitudinal beam polarizations of 50\% on average.
Proton beam bunches with positive and negative helicities at the STAR interaction region were collided in four different helicity combinations, $++$, $+-$, $-+$, and $--$.
STAR events were sorted accordingly. 

\begin{figure}
  \includegraphics[height=.25\textheight]{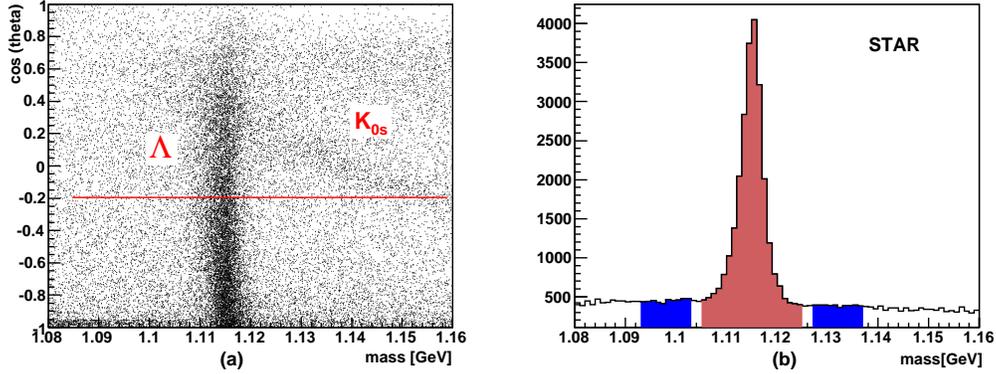}
  \caption{(a) The invariant mass versus $\cos \theta$ for the $\Lambda$ candidates after particle identification and topological selections, and (b) the invariant mass distribution of $\Lambda$ candidates after the additional requirement of $\cos\theta < -0.2$. The invariant mass range for the signals was chosen to be 1115$\pm$10 MeV, and the shaded areas on either side of the signal peak were used in estimating residual background.}
\label{mass}
\end{figure}

The \lll\ was reconstructed via the decay $\Lambda \to p \pi^-$ ($\bar \Lambda \to \bar p \pi^+$) with a branching ratio of $63.9 \pm 0.5\%$~\cite{PDG}.
The \lll\ candidates were reconstructed by pairing proton and pion tracks after their particle identification from specific energy loss $dE/dx$ in the STAR Time Projection Chamber (TPC).
Topological cuts on the decay vertex were applied to further reduce background\cite{mheinz}.
Fig.\ref{mass}(a) shows the reconstructed invariant mass of the $\Lambda$ candidates after these selections versus $\cos \theta$.
The $\Lambda$ signal is seen together with combinatoric background and background originating from $K_s^0 \to \pi^+\pi^-$ decays when the specific energy loss $dE/dx$ is the same for $\pi^+$ and proton.
Candidates with $\cos(\theta) > -0.2$ were excluded from the analysis to eliminate $K_s^0$ background.
This selection reduced the $\Lambda$ signal by 40\%.
Fig.\ref{mass}(b) shows the resulting invariant mass distribution of $\Lambda$.
Candidates outside the invariant mass range $1105 < M < 1125\,\mathrm{MeV}$ were rejected from the analysis.

To extract the \la~ polarization using Eq.(2), the detector acceptance $A(\cos\theta)$ is needed.
It is usually obtained from Monte Carlo simulation.
In this analysis, we use a method to (largely) cancel the detector acceptance and thus do not rely on simulation.
The method uses a physics symmetry of the QCD production, which causes the longitudinal \la~ polarization in singly polarized $pp$ collisions to reverse sign with the helicity of polarized beam.
That is, $P^{+}_\Lambda(\eta)=-P^{-}_\Lambda(\eta)$, where the pseudo-rapidity $\eta$ is defined along the polarized beam and the superscripts + and - refer to the proton beam helicity.
Conceptually, for an experiment with equal luminosities for opposite beam helicities (fully polarized), the asymmetry of $\Lambda$ yields within a small interval in $\cos\theta \in [\cos \theta_1,\cos \theta_2]$ for opposite beam helicities,
\begin{equation}
 \frac{N^+-N^{-}} {N^++N^-} 
\simeq \alpha P_\Lambda  \frac {\cos \theta_1 +\cos \theta_2}{2},
\label{pol}
\end{equation}
is proportional to the polarization $P^+_\Lambda \equiv P_\Lambda$, as seen by substitution of Eq.~\ref{ideal}.
The detector acceptance is canceled.

In a more realistic case, the beam is not fully polarized and the integrated luminosities differ for positively and negatively polarized beams. 
Accordingly, $D_{LL}$ is extracted as follows: 
\begin{equation}
D_{LL}=\frac{1}{\alpha P_\mathrm{beam} <\cos \theta>} \frac{N^+ -R N^-} {N^+ + R N^-},
\label{eq_dll}
\end{equation} 
where $R=L^+/L^-$ is the ratio of luminosities for positively and negatively polarized beams.
At RHIC both beams are polarized and the yields $N^+$ ($N^-$) are thus formed from the yields $n^{++}$, $n^{+-}$, $n^{+-}$, and $n^{-+}$ by beam helicity configuration weighted with the respective luminosities \cite{BBC}. 
The degree of beam polarization $P_{beam}$ is measured with polarimeters \cite{polmeter} for each of the proton beams individually at RHIC.

We have extracted $D_{LL}$ from the spin transfer for \lll\ candidates by subtracting the spin transfer for residual background as follows,
\begin{equation}
D_{LL}=\frac{D_{LL}^\mathrm{raw}-rD_{LL}^\mathrm{bg}}{1-r},
\label{eq_sig}
\end{equation}
where $D_{LL}^{raw}$ is the spin transfer for the \lll\ candidates under the mass peak in Fig.1(b),  $D_{LL}^{bg}$ is the spin transfer for residual background determined from the candidate yields with displaced invariant mass values as indicated in Fig.1(b), and $r$ is the fractional background under the mass peak.

\begin{figure}
  \includegraphics[height=.25\textheight]{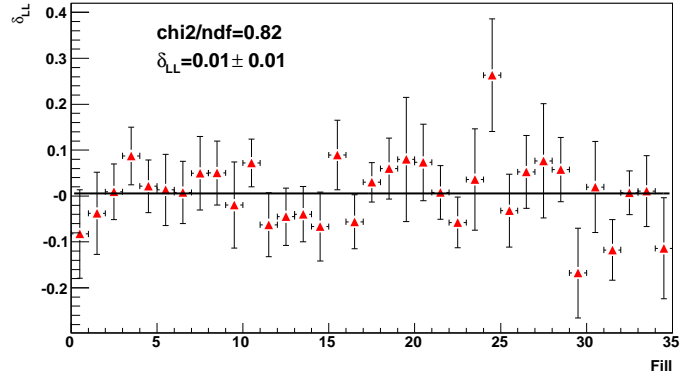}
\caption{Longitudinal ``spin transfer'' $\delta_{LL}$ for inclusive $K_s^0$ versus RHIC beam fill in $pp$ collisions at $\sqrt s=200$ GeV. The indicated uncertainties are statistical only.}
\label{dllk0s}
\end{figure}

As a cross check, we have used the above method for spin-less $K_s^0$ particles to extract the zero-"spin transfer", which will be denoted with $\delta_{LL}$ here.
The $K_s^0 \rightarrow \pi^+ \pi^-$ reconstruction is similar to that for the \lll~. 
The yield is twice larger.
Fig.\ref{dllk0s} shows the results versus RHIC beam fill (time).
The results for $\delta_{LL}$ are consistent with zero to within a statistical error of 0.01, as expected.
Hence, no evidence is found for systematics in the analysis method at the level of 0.01 or larger.

\begin{figure}
  \includegraphics[height=.25\textheight]{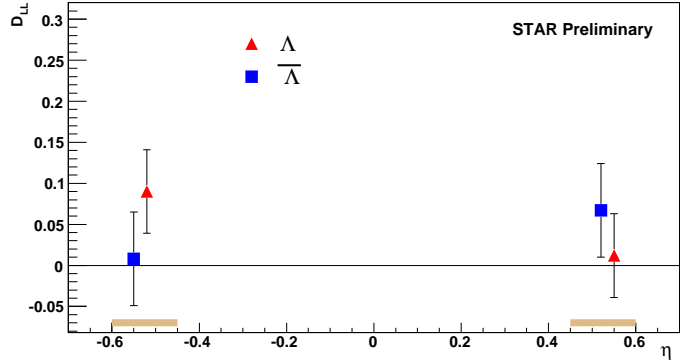}
\caption{Preliminary results for longitudinal spin transfer of $\Lambda$ and $\bar \Lambda$ in $pp$ collisions at $\sqrt s=200$ GeV for positive and negative pseudo-rapidities $\eta$.
The indicated uncertainties on the data points are statistical only.  The shaded band indicates the systematic uncertainty in the STAR measurement.  An additional 20\% scale uncertainty is estimated from uncertainty in the RHIC beam polarization measurement.}
\label{dll}
\end{figure}

After data selections approximately 30K \la's and 24K \al's remain.
Their mean $x_F$ is about 0.0075 and the mean $p_T$ is 1.3 GeV/c.  
Fig.\ref{dll} shows the preliminary results for $D_{LL}$ versus $\eta$. 
Systematic uncertainty at the level of 0.01 is estimated from uncertainties in the relative luminosity ratio $R$ (0.004), non-longitudinal components of beam polarization (2\%), and the decay parameter $\alpha$ (2\%).
The scale uncertainty caused by the online RHIC beam polarization measurement is estimated to be no larger than 20\%.

STAR has measured the $\Lambda+\bar \Lambda$ yield for transverse momenta $1 < p_T < 5\,\mathrm{GeV/c}$ in proton collisions at $\sqrt{s} = 200\,\mathrm{GeV}$~\cite{mheinz}.
The data are reasonably well described by pQCD evaluation with a suitable choice for the fragmentation function.
The present measurement of $D_{LL}$ at small $x_F$ and $p_T$ forms a proof-of-principle measurement that paves the way for measurements at larger transverse momenta, which may give new constraints on the polarized (anti-)strange sea quark distribution in the polarized nucleon.

The author is supported by the U.S. Department of Energy under Contract No. DE-AC02-05CH11231.

\end{document}